\documentclass[11pt]{article}
\usepackage{ascmac,amsmath}
\usepackage{bm}
\usepackage{graphicx,epsfig}
\usepackage{geometry}
\makeatletter
\renewenvironment{thebibliography}[1]
{\section*{\refname\@mkboth{\refname}{\refname}}%
  \list{\@biblabel{\@arabic\c@enumiv}}%
       {\settowidth\labelwidth{\@biblabel{#1}}%
        \leftmargin\labelwidth
        \advance\leftmargin\labelsep
 \setlength\baselineskip{11pt}%
        \@openbib@code
        \usecounter{enumiv}%
        \let\p@enumiv\@empty
        \renewcommand\theenumiv{\@arabic\c@enumiv}}%
  \sloppy
  \clubpenalty4000
  \@clubpenalty\clubpenalty
  \widowpenalty4000%
  \sfcode`\.\@m}
 {\def\@noitemerr
 {\@latex@warning{Empty `thebibliography' environment}}%
\endlist}
\makeatother
\begin{document}
\centerline{{\sl Genshikaku Kenkyu Suppl.} No. 000 (2012)}
\begin{center} 
\vskip 2mm
{\Large\bf


 Measurements of the photon induced production of $\Lambda$ in the $^2$H$({\gamma},{\Lambda})$X process at threshold energies.

\hspace{-1mm}\footnote{Presented at the International Workshop on Strangeness 
Nuclear Physics (SNP12), August 27 - 29, 2012, \\
\hspace*{5mm} Neyagawa, Osaka, Japan.}
}\vspace{5mm}

{
B. Beckford$^a$,  P. Byd$\breve{z}$ovsk$\acute{y}$$^{d}$, A. Chiba$^a$, D. Doi$^a$, T. Fujii$^a$, Y. Fujii$^a$, K. Futatsukawa$^a$,\\ T. Gogami$^a$, O. Hashimoto$^a$,  Y.C. Han$^c$ ,K. Hirose$^b$, S. Hirose $^a$,  R. Honda$^a$, K. Hosomi$^a$, \\T. Ishikawa$^b$,  H. Kanda$^a$, M. Kaneta$^a$, Y. Kaneko$^a$, S. Kato$^a$, D. Kawama$^a$,  C. Kimura$^a$\\ S. Kiyokawa$^a$, T. Koike$^a$,  K. Maeda$^a$, K. Makabe$^a$, M. Matsubara$^a$, K. Miwa$^a$, S. Nagao$^a$,\\ S. N. Nakamura$^a$, A. Okuyama$^a$, K. Shirotori$^a$,  K. Sugihara$^a$, K. Suzuki$^b$, T. Tamae$^b$, H. Tamura$^a$,\\ K. Tsukada$^a$, K. Yagi$^a$,  F. Yamamoto$^a$, T. O. Yamamotoi$^a$, H. Yamazaki$^a$, and Y. Yonemoto$^a$
}\bigskip

{\small
$^a$Department of Physics, Tohoku University, 
Sendai 9808578, Japan \\ 
$^b$Research Center for Electron Photon Science, Tohoku University, Sendai,
982-0826\\ 
$^c$School f Nuclear Science and Technology, Lanzhou University, Lanzhou,
730000 \\ 
$^d$Nuclear Physics Institute, $\breve{R}$e$\breve{z}$ near Prague, Czech Republic,
25068 \\

}
\end{center}
\vspace{3mm}

\noindent
{\small \textbf{Abstract}:\quad
An experiment  was carried out with the NKS2+ in 2010 at the Research Center for Electron Photon Science (ELPH), in which tagged photon beams in the range of 0.8 $\le$ $E_{\gamma}$ $\le$ 1.1 GeV were impinged on a liquid $^2$H target positioned at the center of the NKS2+. The produced $\Lambda$ was subsequently detected by the $p{\pi^{-}}$ decay channel. 
Integrated cross sections of the $^2$H$({\gamma},{\Lambda})$X in the angular region of 0.9 $\le$ $\cos{\theta}_{\Lambda}^{LAB}$ $\le$ 1.0 was derived and compared with preceding experimental results of the NKS2 collaboration.  In addition, the momentum spectra for two photon energy regions were also procured.  The ${\Lambda}$ angle dependent cross sections as a function of the scattering angle in the laboratory system was additionally deduced. We present the latest results on the excitation function of ${\Lambda}$ photoproduction, the momentum distributions, and polarization.
}%


\section{Introduction}
The photoinduced  production of strangeness has been tirelessly investigated on the proton and most recently the emphasis has veered to the neutron to actualize a comprehensive model capable of describing the reaction mechanism. To facilitate this a deuteron target is used to provide a loosely bound neutron target. Our focus is directed in the threshold energy region, in which it is conceded that the reaction will be significantly less disturbed by higher nucleon resonances, allowing for a simplification in the description of the reaction and thus permitting the study with less uncertainty. The sizable experimental data sets that have been measured for the $^1$H$({\gamma},K^{+}) {\Lambda}$ reaction at facilities such as CLAS and SAPHIR, are not adequate to constrain theoretical models and successfully predict the cross section of the unmeasured channels. The goal of this work is to further investigate strangeness photoproduction in the threshold region by focusing on $\Lambda$ production.

\section{The NKS2+ Spectrometer }
The upgraded large acceptance magnetic spectrometer, NKS2+, is composed of a cryogenic target system located in the center of various detector systems that works in tandem with a photon tagging array.  Using a carbon wire as a target, photon beams are produced via the electromagnetic interaction with the electrons orbiting in the Stretcher Booster Ring (STB) in the form of bremsstrahlung radiation. The STB ring is capable of accelerating electrons injected from a 0.15 GeV Linac injector up to 1.2 GeV. 
The photon beam is guided  through a collimator in order to reduce the beam halo, a sweep magnet, and into the NKS2+ and is bombarded on the target. Moving from the inner most position outwards, is the target which is surrounded by a Vertex Drift Chamber (VDC) and an Inner Hodoscope (IH), which acts as the start trigger for time of flight measurements. These detectors are enclosed in a Cylindrical Drift Chamber (CDC). Substantial effort was placed on advancing the study on the neutron as a consequence of the favorable findings of previous experiments with the Neutral Kaon Spectrometer (NKS) and NKS2~\cite{takahashi,watanabe,Kanda,tsukada}; The spectrometer was upgraded by the replacement of the inner detectors in order to further improve upon the acceptance. An experiment has been designed and performed by the NKS2 collaboration was taken at ELPH in 2010.

  \begin{figure}[ht]
  	  \vspace{-.6cm}
    \begin{center}
     \includegraphics[width=10.0cm]{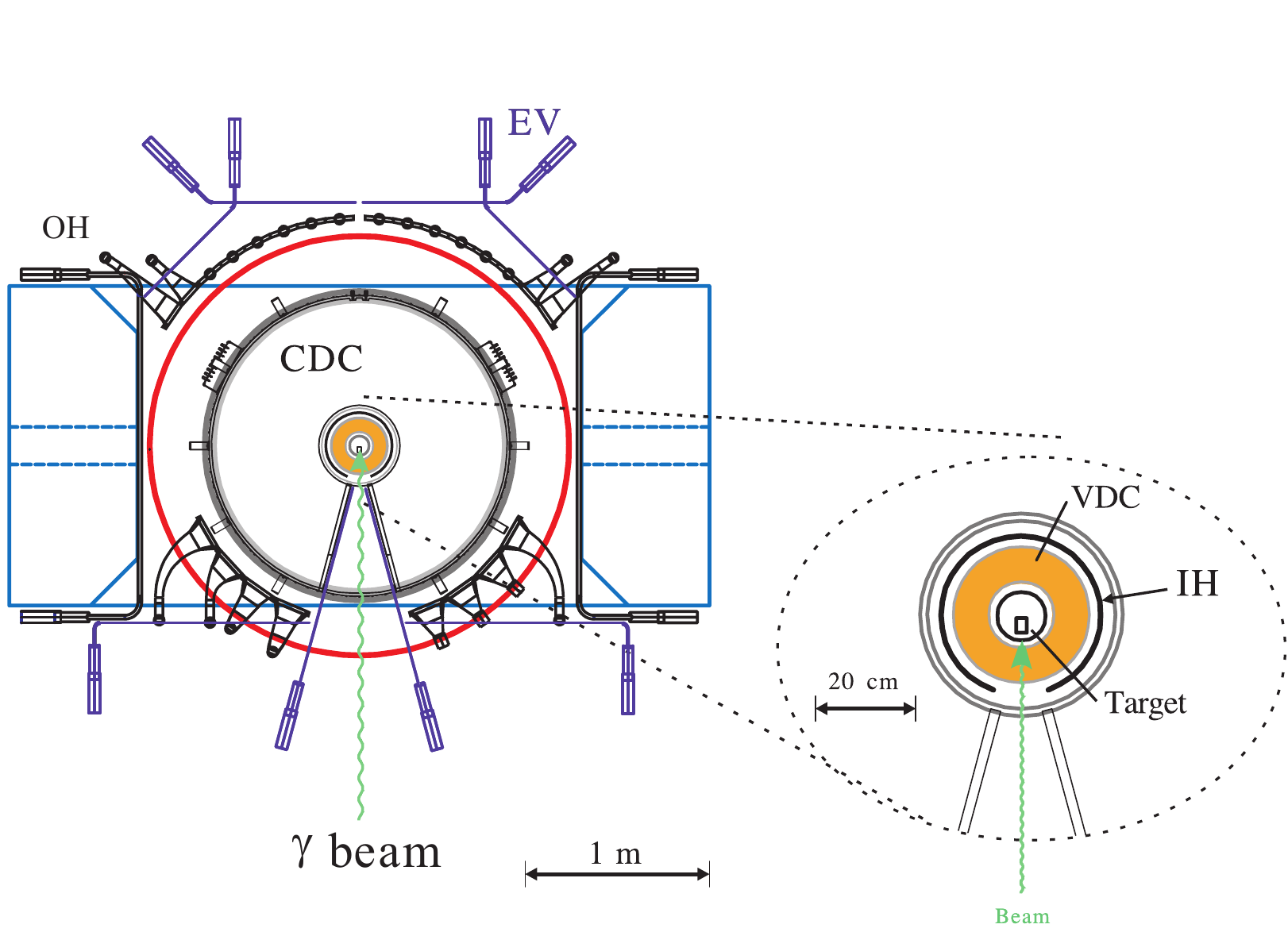}
       \vspace{-.4cm}
      \caption{
        The figure shows a schematic perspective the of recently improved NKS2 spectrometer, henceforth referred to as NKS2+.  The inner detector package has been  redesigned and installed and is visible in the zoomed image on the right. 
      }
      	  \vspace{-.9cm}
      \label{fig:NKS2_upgraded_detectors}
    \end{center}
  \end{figure}
  
  \begin{figure}[h!]
\begin{center}
	      \includegraphics[width=7.15cm]{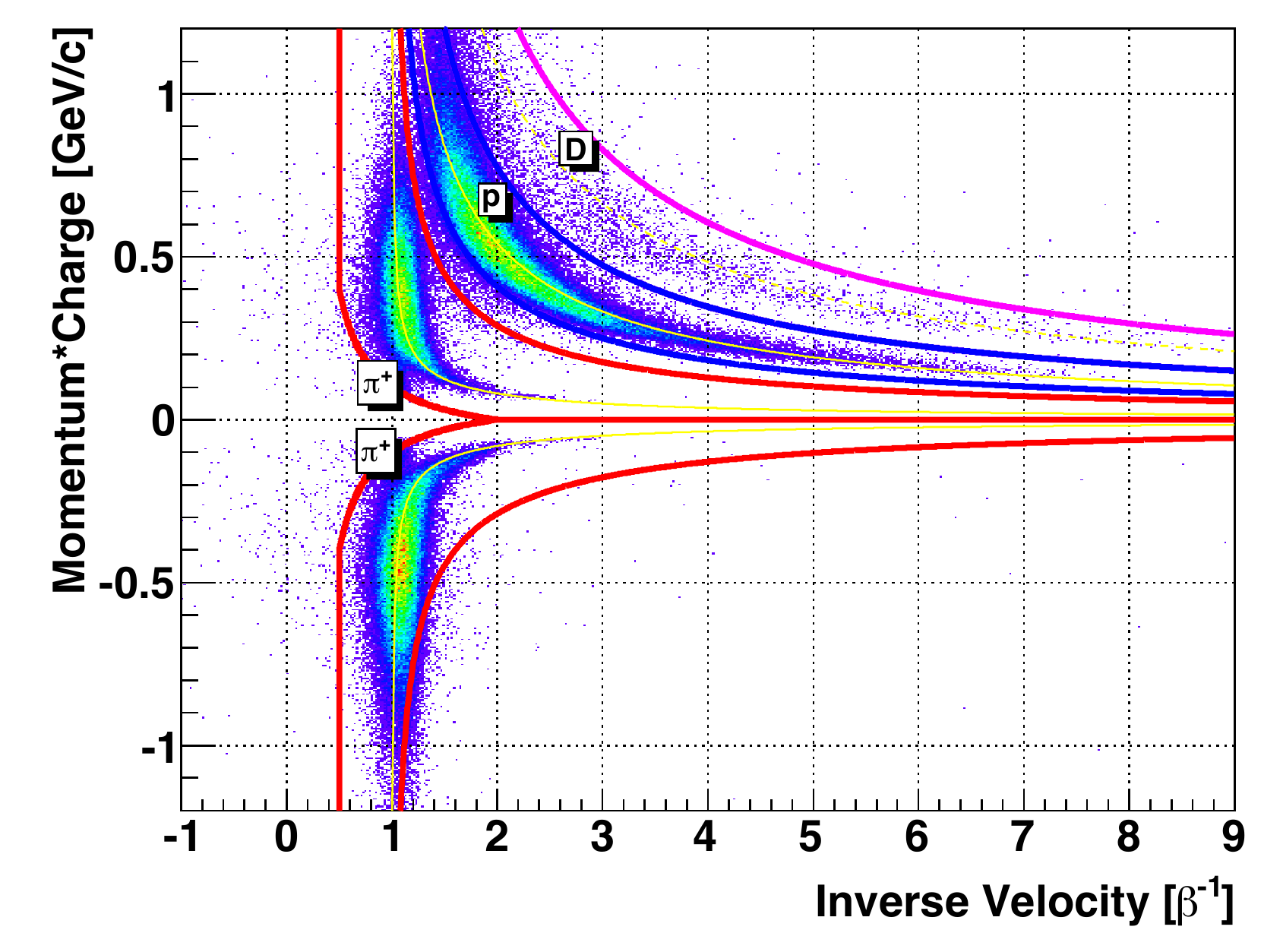}
	      	\includegraphics[width=8.75cm]{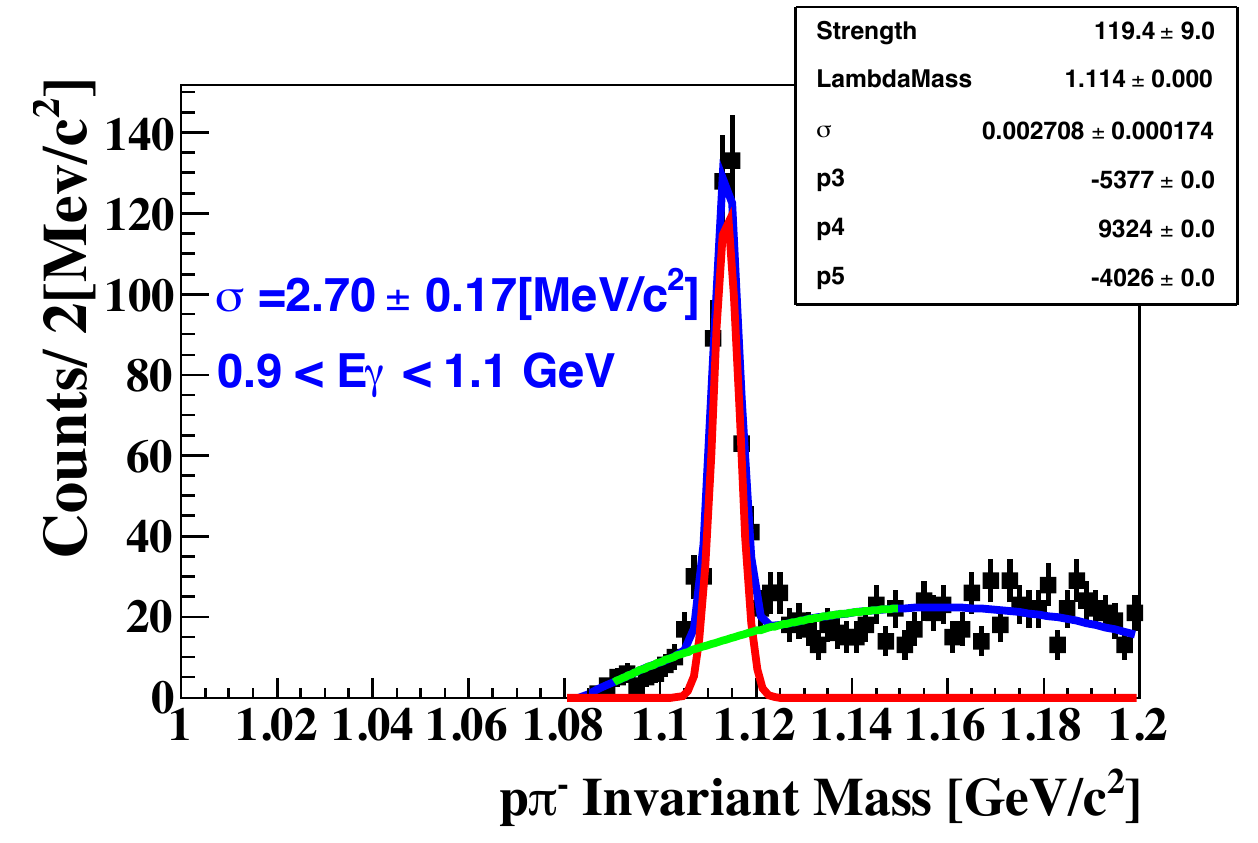}
     	 	\caption{The particle momentum multiplied with the particle charge plotted as a function of the inverse velocity is shown on the left.  The proton and pion selection regions are shown n the curved bands. The charged pions are identified by the sign of the momentum. The $p{\pi^{-}}$ invariant mass resolution for the energy bin of 0.9 $\le E_\gamma \le$1.1 GeV (right) was found by fitting as 2.87 $\pm$ 0.19 MeV/c in rms.  }
     	 \label{fig:momentum_versus_inverse_beta}
 \vspace{-.6cm}
\end{center}
\end{figure}

  \section{Data Analysis}
The experiment was performed with tagged photon beams in the range of 0.8 $\le$ $E_{\gamma}$ $\le$ 1.1 GeV, at tagging rate of $1.5 - 2.5$ MHz. The accepted number of events and recorded photons were 0.64 $\times$ 10$^{9} $ and  0.89 $\times$ 10$^{12}$ respectively. The direct approach to measuring the photo production of strangeness is by reconstructing the invariant mass of the produced particles that contain a strange quark. The momentum determination and tracking of the detected particles was performed by the CDC and the particle species was identified by Time Of Flight (TOF) measurements. . The PID capabilities of the NKS2+ is showed in Figure~\ref{fig:momentum_versus_inverse_beta}, where the particle's momentum multiplied with the charge plotted as a function of the inverse velocity is presented. Additionally, an opening angle selection requirement of $-09$ $\le$ cos${\theta}$ $\le$ 0.9 was used in order to reduce the $e^{+}e^{-}$ contamination.  The produced $\Lambda$ was detected in the $p{\pi^{-}}$ decay channel.
Numerous cuts were applied to the obtained raw invariant mass spectrum to resolve and extract a clean $\Lambda$ distribution from the continuum shown on the right in Figure~\ref{fig:momentum_versus_inverse_beta}.

\section{Results and Discussion}
\begin{figure}[ht]
   	 \begin{center}
	 \includegraphics[width=8.0cm]{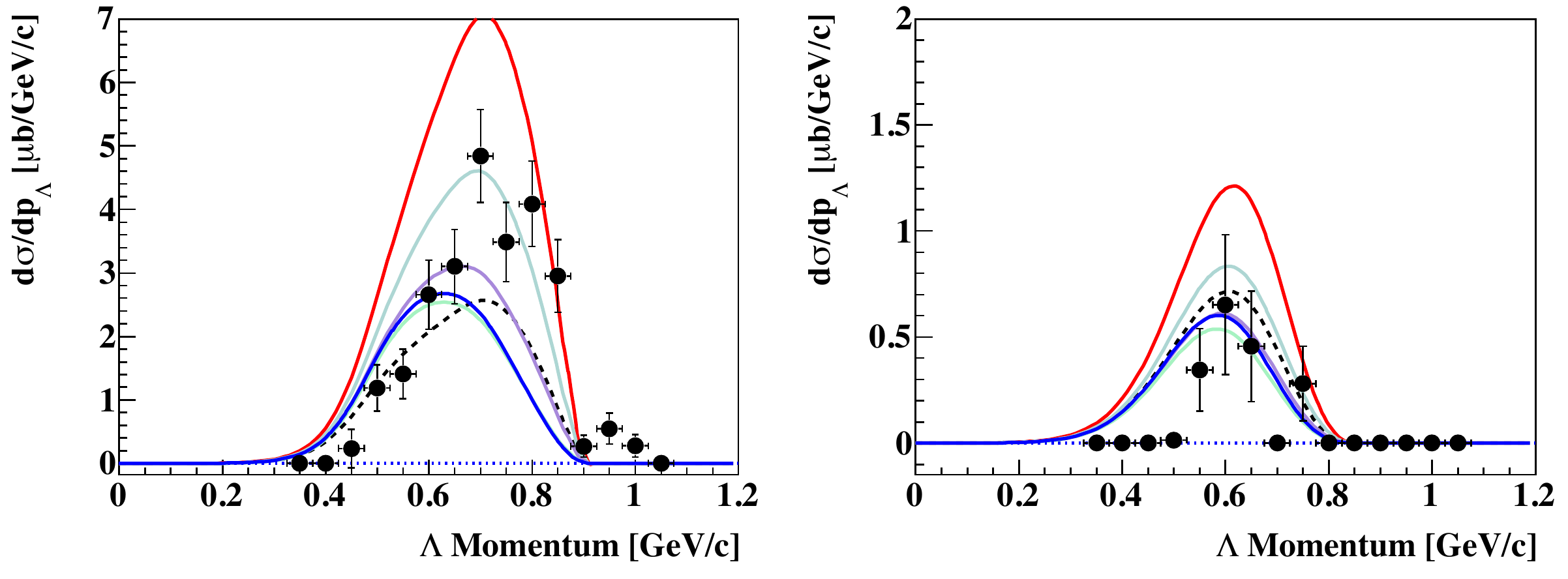}
	  \includegraphics[width=8.0cm]{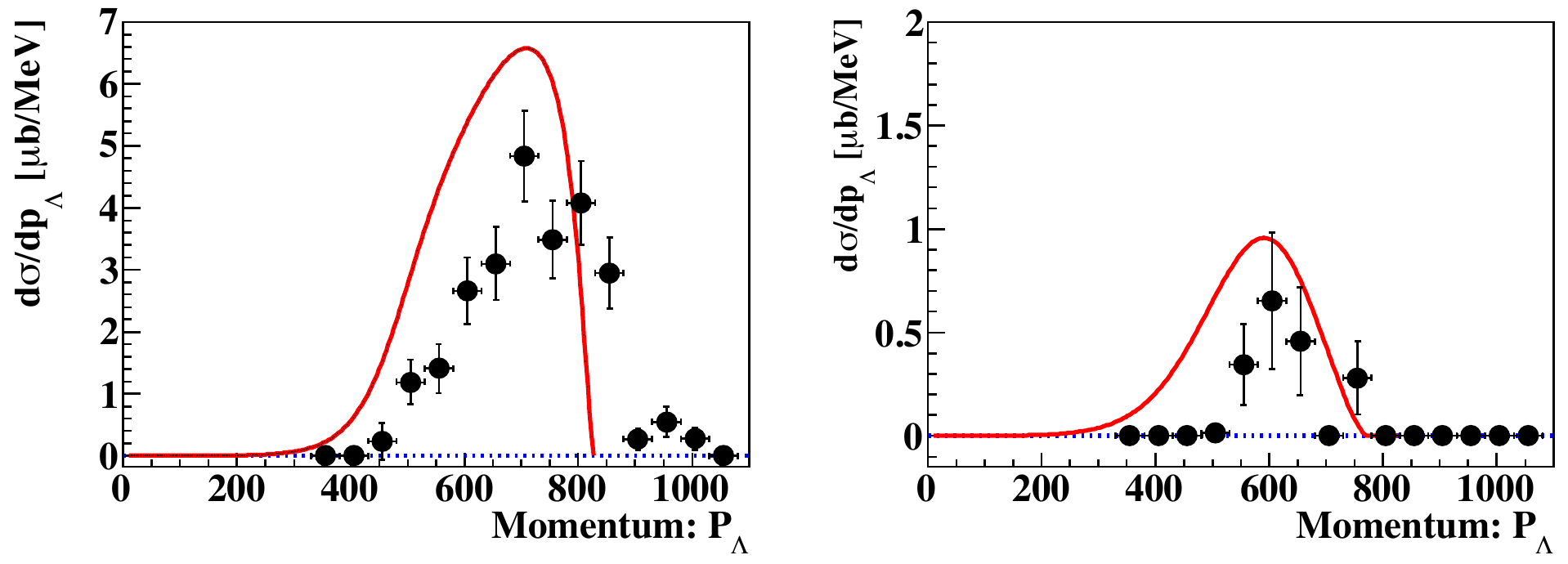}
	 \includegraphics[width=8.2cm]{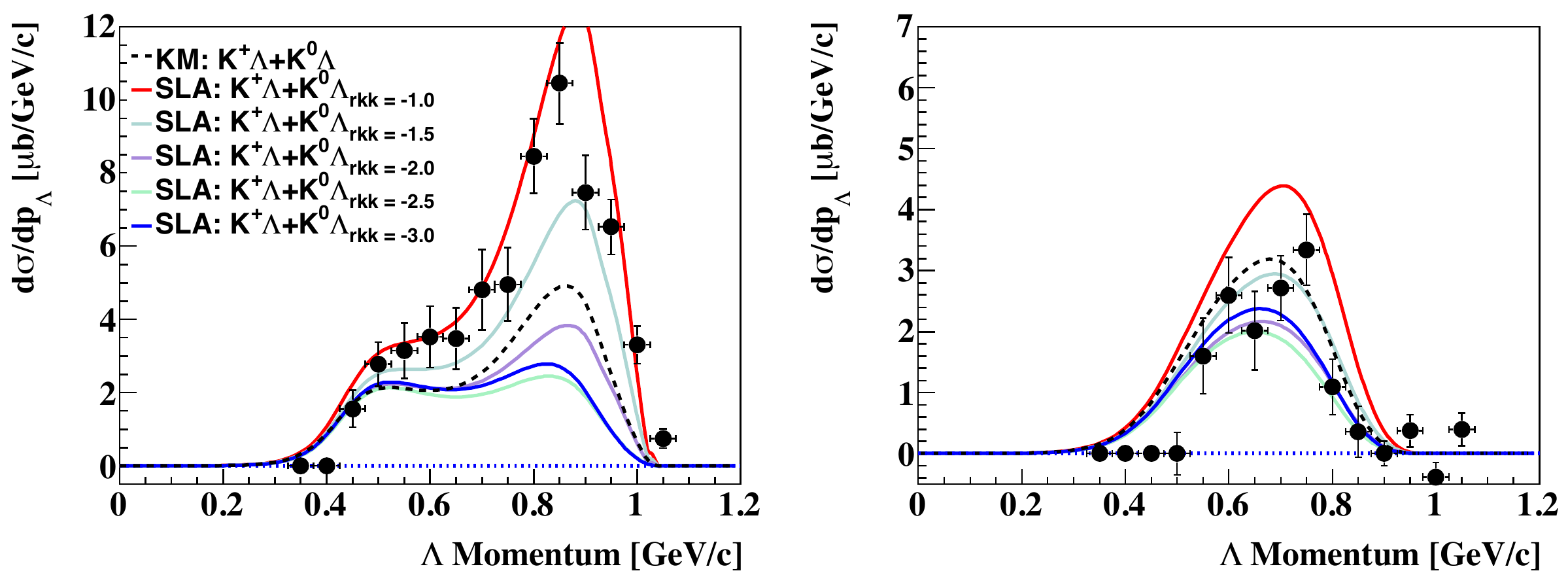}
         \includegraphics[width=8.2cm]{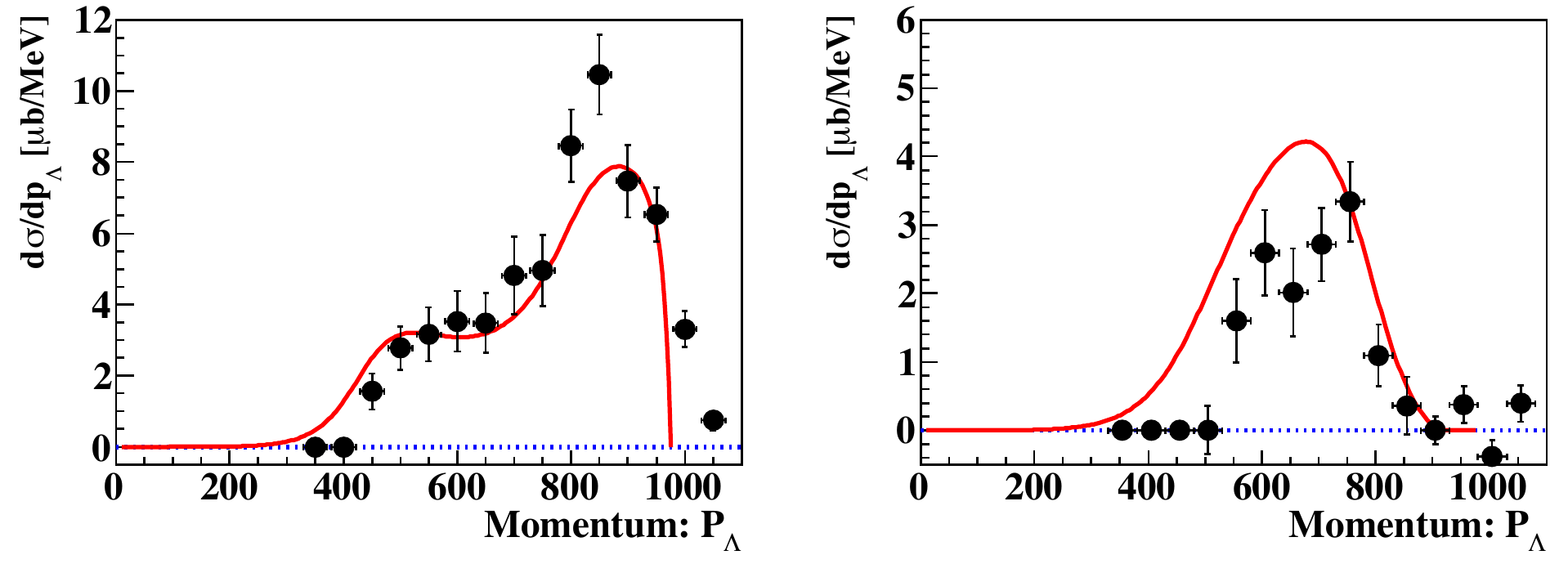}
	  \vspace{-.6cm}
      \caption{
Comparison of results with KM and SLA calculations of the semi-inclusive ${\Lambda}$ momentum cross sections for the addition of contributions of the $^2$H$({\gamma},K^{+}) {\Lambda}n$ and $^2$H$({\gamma},K^{0}) {\Lambda}p$ reactions  for photon beam energy of  (top) $0.9 - 1.0$ and (bottom) $1.0 - 1.08$ GeV in the laboratory frame.  The angular integration regions of $0.95 - 1.0$ and $0.90 - 0.95$ (cos$\theta_{\Lambda}^{LAB}$) are shown left to right respectively. The $r_{K_{1}K\gamma}$ parameter for the SLA models is displayed in the figure. The RPR comparisons are given in the right two columns. }
      \label{fig:Comparison_SLA_KM_Lambda_singles}
    \end{center}
  \end{figure}

We next proceed to employ various theoretical calculations as a comparison medium for the experimentally yielded outcomes. This initial comparison is made with the momentum dependent differential cross section obtained for the inclusive measurement of ${\Lambda}$ to the Kaon-MAID (KM), Saclay-Lyon A (SLA) and Regge-plus-resonance (RPR-2007) models. The comparison of results with KM and SLA calculations for the addition of amplitudes of the $^2$H$({\gamma},K^{+}) {\Lambda}n$ and $^2$H$({\gamma},K^{0}) {\Lambda}p$ reactions for incident photon beam energy of (top) $0.9 - 1.0$ and (bottom) $1.0 - 1.08$ GeV in the laboratory frame.  The angular regions of  $0.95 - 1.0$ and $0.90 - 0.95$ (cos$\theta_{\Lambda}^{LAB}$)  are shown left to right respectively in the left two columns of Figure~\ref{fig:Comparison_SLA_KM_Lambda_singles}. The $r_{K_{1}K\gamma}$ coupling constant, which is a parameter for the SLA model, is displayed in the figure.  The RPR-2007 comparisons are seen in the right two columns. SLA $r_{K_{1}K\gamma}$ = $-(1.0 -1.5)$ and RPR-2007 adequately reproduce the size and shape of the data, especially in the higher energy bin.

\begin{figure}[h!]
\begin{center}
\vspace{-.5cm}
        \includegraphics[width=8.15cm]{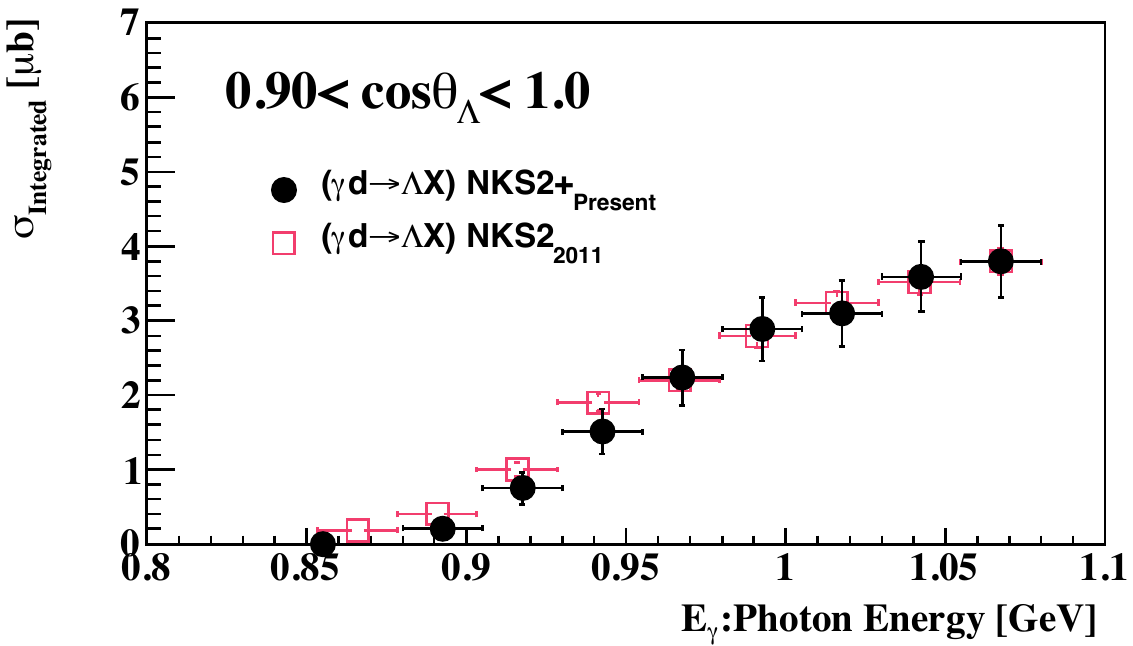}
        \includegraphics[width=8.15cm]{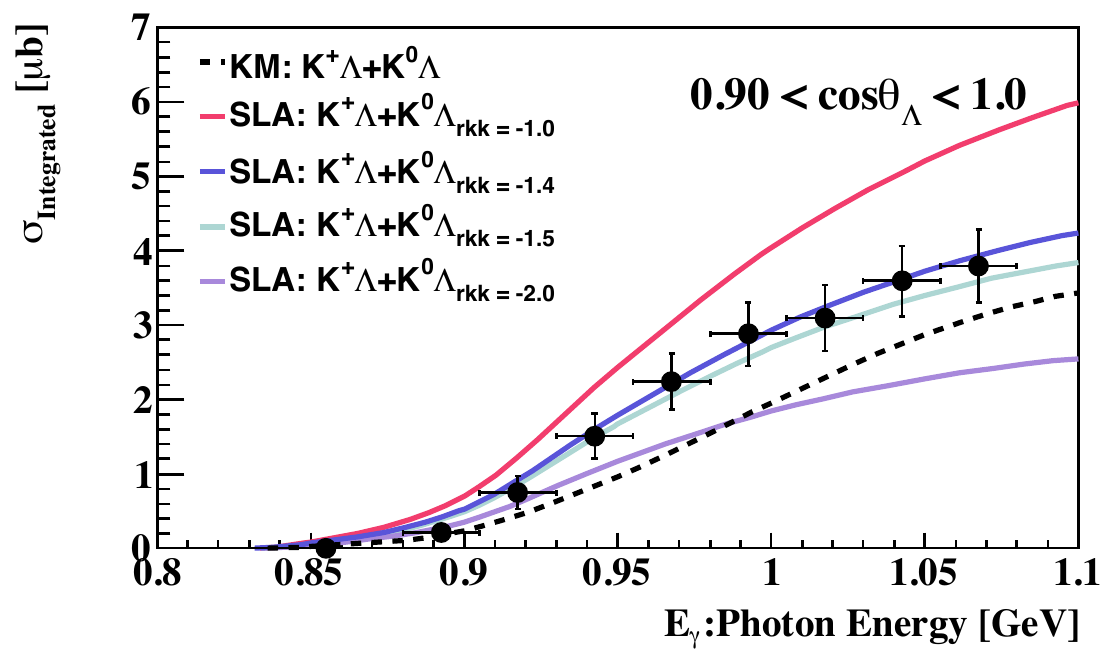}
	\caption{The results of the NKS2~\cite{futatsukawa} measurements of the energy dependent integrated cross section are shown with the NKS2+ results as open squares and black circles respectively (left). Comparison to integrated cross section as a function photon energy predicted by the SLA and KM theoretical models (right).  Where the cross sections of $^2$H$({\gamma},K^{+}) {\Lambda}n$ and $^2$H$({\gamma},K^{0}) {\Lambda}p$ reactions are summed.  }
\vspace{-.3cm}
 \label{fig:total_gd_KpK0_L_SLA_comparison}
\end{center}
\end{figure}

The experimental results of the inclusive measurement of ${\Lambda}$ on a deuterium target using the NKS2 spectrometer prior to the recent upgrade was successfully carried out and the results were reported~\cite{futatsukawa,Beckford_Baryon10}. 
The background subtracted energy dependent cross section of the inclusive measurement of ${\Lambda}$ photoproduction in $^2$H$({\gamma},{\Lambda})$X reaction for the angular integrated region of 0.9$\le$cos${\theta}_{\Lambda}^{LAB}$ $\le$1.0 for the NKS2~\cite{futatsukawa,Beckford_Baryon10} and the latest results, obtained from independent experiments and analysis, are presented in Figure~\ref{fig:total_gd_KpK0_L_SLA_comparison} as the solid circles and open squares respectively. Only statistical errors are shown. The latest results furnished an excitation function that has considerable agreement with the finalized results of the preceeding experimental, emphasizing the uniqueness of the research, and more importantly, the reliability of the data reported by the NKS2/NKS2+ collaboration.
The SLA predictions describe well the data for $r_{K_{1}K\gamma}$ = $-1.4$ shown on the right in Figure~\ref{fig:total_gd_KpK0_L_SLA_comparison}.

We next report on our measurements of the ${\Lambda}$ recoil polarization observables.  
The self analyzing nature of the weak decay of the  ${\Lambda}$ hyperon gives rise to not only parity violation but also an asymmetry. 
The polarization,${P_{\Lambda}}$, was computed from the decay asymmetry of the angular distributions and is calculated by Equation~\ref{eq:lambda_polarization}.

  \begin{equation}
	{P_{\Lambda}}  = \frac{2}{{\alpha} }\frac{N_1 - N_2} {N_1 + N_2 }
		\label{eq:lambda_polarization}
  \end{equation}
  
Where ${\alpha}$ = 0.6421 $\pm$ 0.013, is the decay asymmetry parameter, $N_1$ and $N_2$ are the total number of events where cos${\theta}$ is greater or less than zero respectively. 
It this framework cos${\theta}$ denotes the angle that lies between the proton from the $\Lambda$ decay and the normal of the reaction plane.
This asymmetry, typically thought of as the ratio between the difference and the addition of the competing patterns is a consequence of the witnessed interference between the \emph{s} and \emph{p} waves.

  \begin{figure}[ht]
   \begin{center}
    \includegraphics[width=5.25cm]{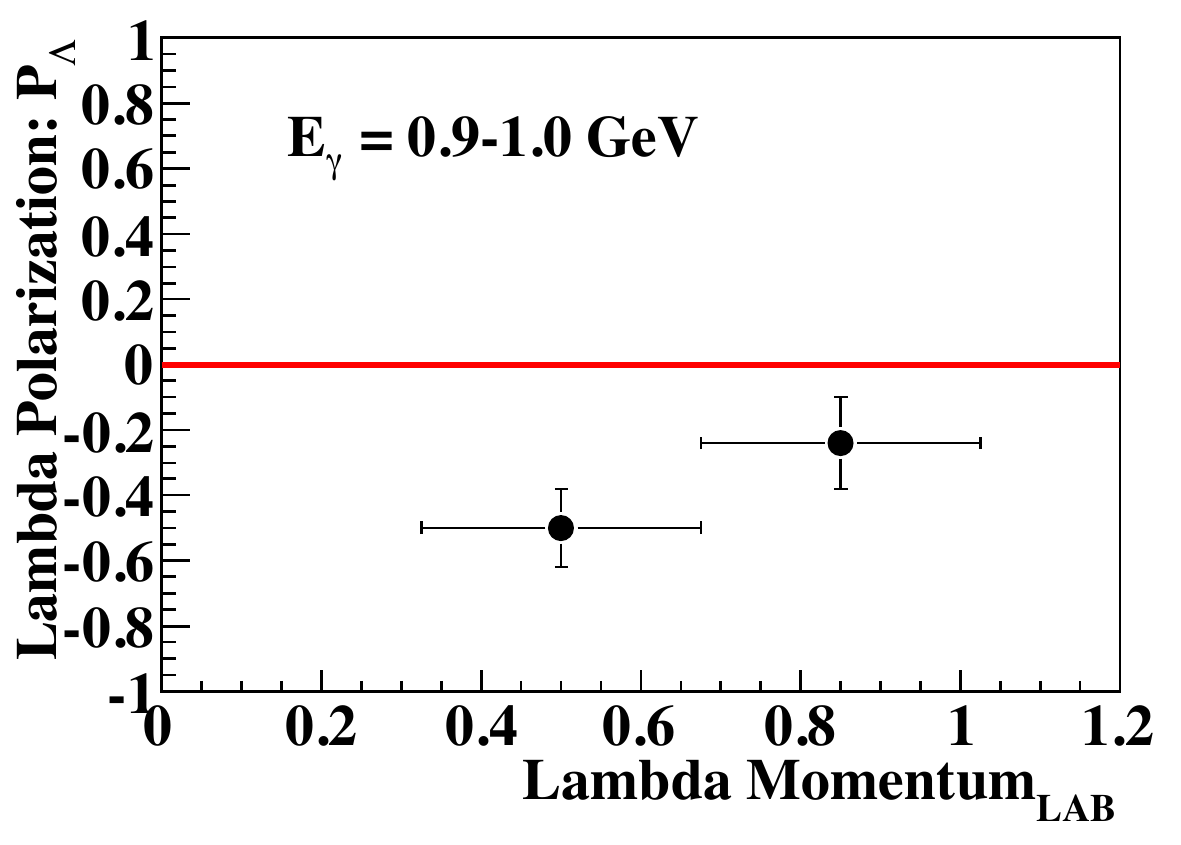}
    \includegraphics[width=5.25cm]{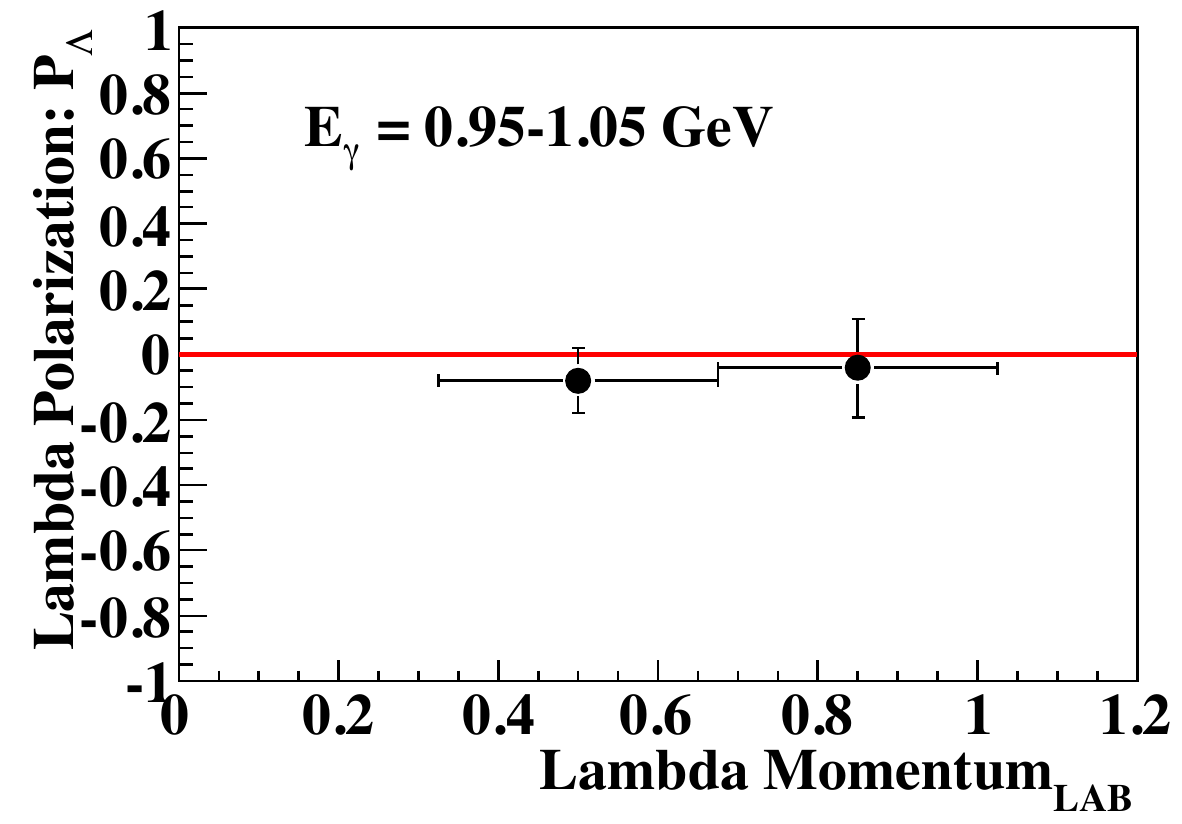}
    \includegraphics[width=5.25cm]{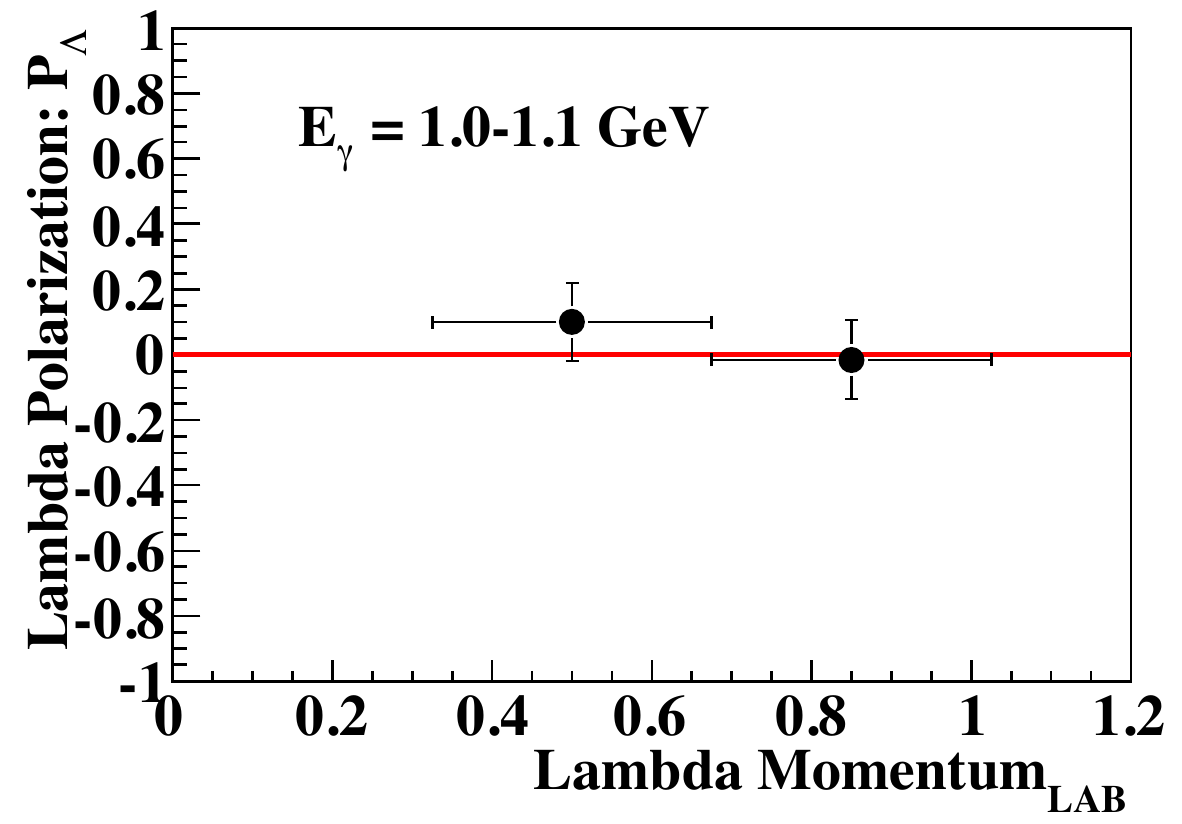}
      \caption{${\Lambda}$ recoil polarization as a function of  momentum in the laboratory frame for the energy range of $0.9 - 1.0, 0.95 - 1.05$,  and $1.0 - 1.1$ GeV. 
      }
        \vspace{-.5cm}
      \label{fig:polarization_of Lambda_mom_E_ranges}
    \end{center}
  \end{figure}

The momentum dependence of ${\Lambda}$ polarization was determined for three photon energy ranges,  $0.9 - 1.0, 0.95 - 1.05$,  and $1.0 - 1.1$ GeV, in the laboratory frame. The results of polarization of the energy ranges are shown in Figure~\ref{fig:polarization_of Lambda_mom_E_ranges},  left to right respectively.  The polarization in the lower energy bin is distinctively 
negative. Whereas, in the other energy bins, the value of the polarization is consistent with zero.

\begin{figure}[h!]
\begin{center}
	\includegraphics[width=8.15cm]{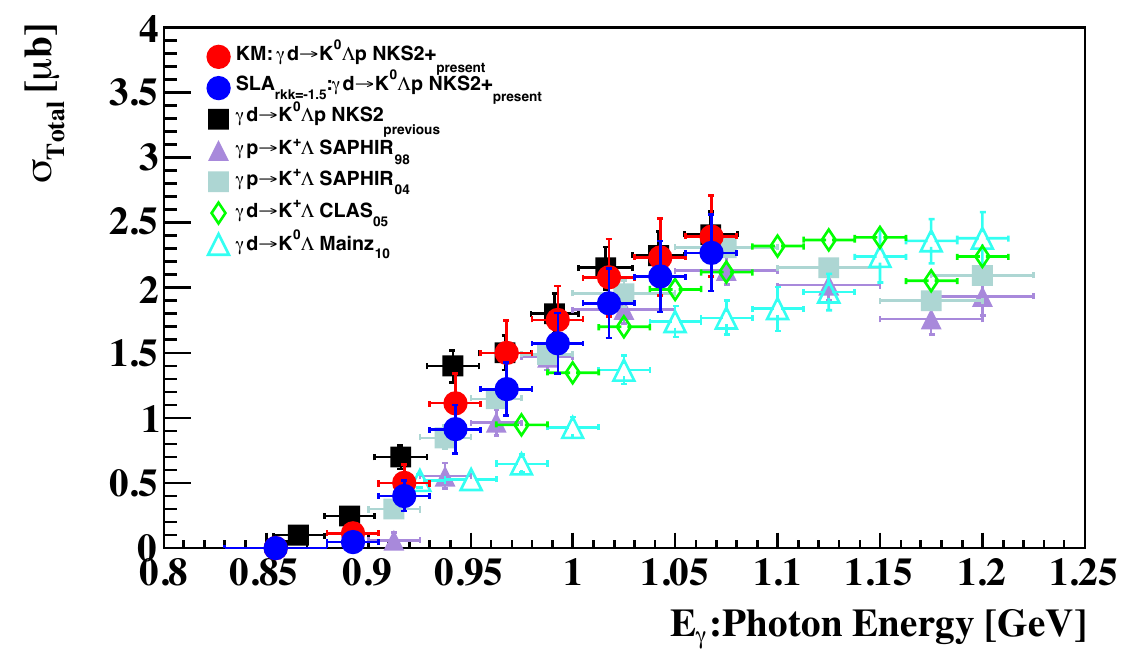}
	\includegraphics[width=8.15cm]{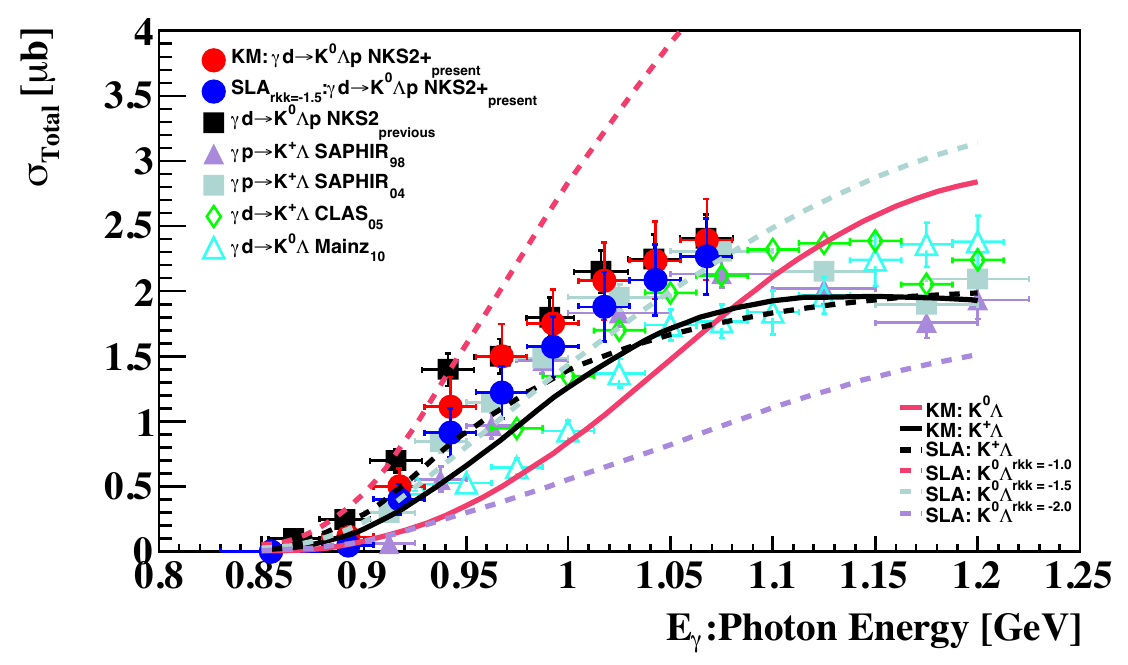}
	\caption{Comparison of the estimated total $^2$H$({\gamma},K^{0}) {\Lambda}p$ cross section with theoretical predictions and published K$^{+}\Lambda$ and K$^{0}\Lambda$ data using proton and deuteron targets~\cite{SAPHIR_data,SAPHIR_data_98,CLAS_data, futatsukawa_dissertation, bantawa_phd}. Only the statistical errors are shown. }
	\label{fig:total_exp_data_comparison_theory}
	   \vspace{-.8cm}
\end{center}
\end{figure}

\begin{equation}
\sigma_{\mathrm {Total}}^{K^{0}\Lambda} =( \sigma_{\mathrm{Integrated}}^{\gamma d\rightarrow \Lambda X} - \tilde{\sigma}_{\mathrm{Integrated}}^{'\gamma d\rightarrow K^{+}\Lambda n}) \cdot \frac{\tilde{\sigma}_{\mathrm{Total}}^{'\gamma d \rightarrow K^{0}\Lambda p} }{\tilde{\sigma}_{\mathrm{Integrated}}^{'\gamma d \rightarrow K^{0}\Lambda p}} 
\label{eq:k0_total_estimation_equation}
\end{equation}

The$^2$H$({\gamma},K^{0}) {\Lambda}p$ total cross section was extrapolated by equation~\ref{eq:k0_total_estimation_equation}.
In the equation, the total and integrated cross of the $\gamma d \rightarrow K^0 \Lambda p$ process are denoted as  $\tilde{\sigma}_{Total}^{'\gamma d \rightarrow K^{0}\Lambda p}$ and $\tilde{\sigma}_{Integrated}^{'\gamma d \rightarrow K^{0}\Lambda p}$. It follows that the integrated cross of the $\gamma d \rightarrow K^+ \Lambda n$ process is designated as $\tilde{\sigma}_{Integrated}^{'\gamma d\rightarrow K^{+}\Lambda n}$. In the above calculation, all integrated cross sections were derived  for the angular integration range of 0.9 $\le$ cos$\theta_{\Lambda}$ $\le$ 1.0 in the laboratory frame of reference.  The values introduced into the calculation for $\tilde{\sigma}_{Total}^{'\gamma d \rightarrow K^{0}\Lambda p}$, $\tilde{\sigma}_{Integrated}^{'\gamma d \rightarrow K^{0}\Lambda p}$ and $ \tilde{\sigma}_{Integrated}^{'\gamma d\rightarrow K^{+}\Lambda n}$ were procured by purely theoretical means. 
The systematic error associated with this method  is approximated at less than 15\%. There is a minor model dependence on the extracted total cross section deduced by using either theoretical calculations. 

Hence, the estimated total cross section derived with the KM and SLA  $r_{K_{1}K\gamma}$= $-1.5$ can be juxtaposed to experimental data. The published results by SAPHIR~\cite{SAPHIR_data, SAPHIR_data_98} and CLAS~\cite{CLAS_data}, and those taken from the dissertations of K. Bantawa\cite{bantawa_phd} and K. Futatsukawa\cite{futatsukawa_dissertation} are plotted alongside the estimated total cross section of $^2$H$({\gamma},K^{0}) {\Lambda}p$ by the method described in equation~\ref{eq:k0_total_estimation_equation} in Figure~\ref{fig:total_exp_data_comparison_theory}.
The maximum photon energy range of the NKS2+ data analyzed in this work is $E_{\gamma}$= 1.08 GeV. While results of the other collaborations extended into a higher energy regimes. The total $^2$H$({\gamma},K^{0}) {\Lambda}p$ compare favorably well to the measured $^1$H$({\gamma},K^{+}) {\Lambda}$ energy dependent cross sections up to 1.08 GeV, this implies that cross sections are roughly on par in terms of overall magnitude. Despite the general agreement between our semi-inclusive energy dependent cross section to those of the SAPHIR and CLAS results, the MAINZ results exhibits a different trend. Our results over the integrated energy range of $0.9 - 1.08$ GeV has a calculated value of $40 - 20$\% higher than the results reported by their group.

Model predictions for the inclusive $^2$H$({\gamma},K^{+}) {\Lambda}n$ and $^2$H$({\gamma},K^{0}) {\Lambda}p$, calculated in the KM and SLA framework, are compared to the experimental results obtained in the threshold energy region gathered by the NKS2, NKS2+, CLAS, SAPHIR,  and MAINZ collaborations. These comparison are shown in Figure~\ref{fig:total_exp_data_comparison_theory} where the curves for the $^2$H$({\gamma},K^{+}) {\Lambda}$n process predicted by KM and SLA and drawn as solid and dashed black lines.  Calculations of the $^2$H$({\gamma},K^{0}) {\Lambda}p$ for KM and SLA,$r_{K_{1}K_\gamma} $ of $-(1.0 - 2.0)$, are shown as the solid light red, dashed light red, dashed, light blue, and dashed light purple curves respectively. The KM  and SLA predictions of the $^2$H$({\gamma},K^{+}) {\Lambda}$n process are reasonably similar and provides an great description of the data up to $E_{\gamma} \le $1.0 GeV, where after the both predictions slightly under shoot the data between 1.0 $E_{\gamma} \le $1.1 GeV and then return to a consistent agreement.  The general consensus between our experimental results and theoretical predictions of the SLA model is good. The shape and size of the extracted total $^2$H$({\gamma},K^{0}) {\Lambda}p$ cross section is acceptably reproduced for the $r_{K_{1}K_\gamma} $ value around $-1.4$.

\section{Acknowledgements} 
 \vspace{-.3cm}
Theoretical calculations for RPR-2007 model were made by P. Vancraeyvald~\cite{RPR_private} of Ghent University.
The research was conducted under the patient and indispensible guidance of the late Professor Osamu Hashimoto.  May he find peace and know that his legacy continues on.

\end{document}